\documentclass{article}
\usepackage {graphicx}
\begin{document}
\baselineskip .75cm 
\begin{titlepage}
\title{\bf Relativistic harmonic oscillator model for quark stars}       
\author{Vishnu M. Bannur  \\
{\it Department of Physics}, \\  
{\it University of Calicut, Kerala-673 635, India.} }   
\maketitle
\begin{abstract}

The relativistic harmonic oscillator (RHO) model of hadrons is used to study quark stars. The mass-radius relationship is obtained and compared with bag model of quark star, using Tolman-Oppenheimer-Volkoff equation. In this model, the outward degenerate pressure due to discrete Landau levels and Landau degeneracy balances the inward gravitational pressure. Where as in bag model the degenerate pressure is due to the standard continuum levels which balances the combined inward pressure due to gravitation and bag pressure. So in RHO model, the confinement effect is included in the degenerate pressure. We found a qualitative similarity, but quantitative differences in mass-radius relationship of quark stars in these two models. Masses and radii are relatively larger and the central energy densities, required for stable quark stars, are lower in RHO model than that of bag model.     
\end{abstract}
\vspace{1cm}
                                                                                
\end{titlepage}
\section{Introduction :}

The relativistic harmonic oscillator (RHO) model of hadrons was extensively studied and successful to explain magnetic moments of baryons \cite{ka.1}, hadron spectroscopy and decays \cite{vi.1}. An attempt was also made to derive it from the first principle in Ref. \cite{vi.2}. It is also used to explain nucleon-nucleon force and study nuclear properties \cite{ka.2}. As a further extension of the model, here we apply it to study quark stars. Quark star may be a big hadron made up of many quarks, subjected to confinement. Such a matter may be found deep inside a neutron star or may be a next fate of a neutron star whenever it's mass exceeds the limit such that the degenerate pressure due to neutrons fail to support the gravitational collapse \cite{gl.1}. Such a compact stars were studied using different confinement models, different equation of state (EoS) and also models with mixture of hadrons and quarks, called hybrid stars \cite{ny.1}. Other exotic quark stars, motivation of these studies, and possible candidates for quark stars are reviewed in Ref. \cite{xu.1}. It is a hope that the study of quark matter in astrophysics may shed light into perturbative and nonperturbative properties of QCD (quantum chromodynamics). It is speculated that certain kind of pulsars may be a possible candidates for quark stars \cite{xu.1}. Our aim here is to study nonrotating, cold, pure quark star confined by RHO potential using Tolman-Oppenheimer-Volkoff (TOV) equation. Of course, such a study was carried out earlier in Ref. \cite{mi.1} where the EoS, needed in TOV equation, was derived by assuming that all quarks are confined to the center of a spherical bag by RHO potential. However, we know, from the study of statistical mechanics of particles in harmonic oscillator potential  \cite{pa.1}, that pressure of such a system goes to zero at infinite volume limit. In our model, we derive EoS by assuming that quarks are confined at different centers in a finite spherical volume. It is similar to the well known problem of diamagntism \cite{pa.1} where electrons are confined at different centers in a plane perpendicular to a uniform magnetic field in a finite volume. This model gives finite pressure at infinite volume limit, as shown in Ref. \cite{pa.1,ba.1}. In Ref. \cite{ba.1}, we also studied an approximate mass-radius relationship of quark stars using Newtonian gravitational theory. 
      
\section{RHO model for quark stars:} 

It is speculated by many authors \cite{ny.1,xu.1} that, before the massive star collapses to black hole when nuclear fuel burns out, it undergoes various stages like white-dwarf, neutron star, hybrid star, quark star or other exotic quark stars. Here we are interested in RHO confinement on bulk quark system like quark stars. It is a simplified model of quark star without rotation and hybrid nature of actual quark star. It is a system of quarks with overall color neutral, but confined by QCD vacuum, like big hadron. There are such models with confinement borrowed from bag models of hadron spectroscopy \cite{ra.1} and studied using both TOV and Newtonian formulations. In bag model of quark star, quarks are free inside the bag, but confined by both bag pressure and gravitational pressure. In our model, quarks are inside a spherical bag, but each quark subjected to a RHO potential with confinement centers anywhere within the bag. This is just the extension of the 2-dimensional confinement of electrons in the problem of diamagnetism to 3-dimension. 

From the hadron spectroscopy studies, the single particle energy eigenvalue in RHO is \cite{ka.1},  
\begin{equation}
\epsilon_n = \sqrt{(2 n+1) C_q^2 + m_q^2} \,\,, 
\end{equation}
where $n = 1,2,...\infty$, $C_q$ is the RHO coupling strength and $m_q$, the quark mass. Therefore, quarks inside a volume $V$ have discrete Landau levels instead of continuum levels as in the case of free quark ($C_q = 0$) inside a bag at infinite volume limit. However, according to Landau's prescription \cite{pa.1}, the continuum energy levels for $C_q = 0$ have become discrete, but degenerate to accommodate all levels of continuum in RHO. This phase-space degeneracy may be written as, 
\begin{equation}
g_n = \frac{V C_q^3}{6 \pi^2} \left[ (2 n +3)^{3/2} -  (2 n +1)^{3/2} \right] \,\,, 
\end{equation}
a straight forward extension of degeneracy in the problem of diamagnetism \cite{pa.1}. 
Now following the standard procedure of evaluation of degenerate pressure, we first evaluate the energy, 
\begin{equation}
E_0 = g_f \sum_{n=1}^{n_F} g_n \epsilon_n = \frac{g_f V C_q^3}{6 \pi^2} \sum_{n=1}^{n_F} \left[ (2 n +3)^{3/2} -  (2 n +1)^{3/2} \right] \sqrt{(2 n+1) C_q^2 + m_q^2} \,\,, 
\end{equation}
and the number of quarks,
\begin{equation}
N_0 = g_f \sum_{n=1}^{n_F} g_n  = \frac{g_f V C_q^3}{6 \pi^2} \sum_{n=1}^{n_F} \left[ (2 n +3)^{3/2} -  (2 n +1)^{3/2}\right]  \,\,, 
\end{equation}
where $g_f$ is the internal degrees of freedom of quarks and $n_F$ is the Fermi level up to which all Landau levels are completely filled. Since each level is degenerate, to take account of partially filled level, expression for energy may be modified as,
\begin{equation}
E_0 = \frac{g_f V C_q^3}{6 \pi^2} \sum_{n=1}^{n_F} \left[ (2 n +3)^{3/2} -  (2 n +1)^{3/2} \right] \sqrt{(2 n+1) C_q^2 + m_q^2} + (N -N_0) \sqrt{(2 n_F+3) C_q^2 + m_q^2}\,\,, 
\end{equation}
where $N$ is the total number of quarks including the partially filled level. The degenerate pressure is $P = - \frac{\partial E_0}{\partial V}$. This degenerate pressure balances the gravitational inward pressure according to the TOV equation,
\begin{equation}
\frac{d\bar{P}}{d\bar{r}} = - \frac{\bar{M}}{\bar{r}^2} \left[ \bar{\varepsilon} + \bar{P}\right] \left[ 1 + \frac{4 \pi \bar{r}^3 \bar{P}}{\bar{M}}\right] \left[ 1 - \frac{2 \bar{M}}{\bar{r}}\right]^{-1} \,\,, \label{eq:tovp} 
\end{equation}
and 
\begin{equation}
\frac{d\bar{M}}{d\bar{r}} = 4 \pi \bar{r}^2 \bar{\varepsilon} \,\,, \label{eq:tovm}   
\end{equation}
where $\varepsilon \equiv \frac{E_0}{V}$, energy density, and $M$ is the mass of quark star. All variables above are normalized as, 
\begin{equation}
\bar{P} = \frac{6 \pi^2}{g_f C_q^4} P\,\,;\,\, \bar{\varepsilon} = \frac{6 \pi^2}{g_f C_q^4} \varepsilon\,\,; \,\,\bar{M} = \left( \frac{g_f}{6 \pi^2}\right) ^{1/2} C_q^2 G^{3/2} M\,\,;\,\,\bar{r} = \left( \frac{g_f}{6 \pi^2}\right) ^{1/2} C_q^2 G^{1/2} r\,\,. 
\end{equation}
As usual, TOV equations are solved, numerically, starting from initial value $n_F$, which corresponds to some central density, and the radius for which pressure goes to zero is obtained. This is repeated for different central densities to obtain the mass-radius relationship, the maximum mass, $\bar{M}_{max}$, of the star and corresponding radius $\bar{R}$ are obtained. The actual or unnormalized  mass and radius are obtained from 
\begin{equation}
M = \bar{M} \left( \frac{6 \pi^2}{g_f}\right)^{1/2} \frac{1}{C_q^2 G^{3/2}} \,\,
\end{equation}
\begin{equation}
R = \bar{R} \left( \frac{6 \pi^2}{g_f}\right)^{1/2} \frac{1}{C_q^2 G^{1/2}} \,\,, 
\end{equation}
and plotted in Fig. 1. 

\begin{figure}[h]
\centering
\includegraphics[height=8cm,width=12cm]{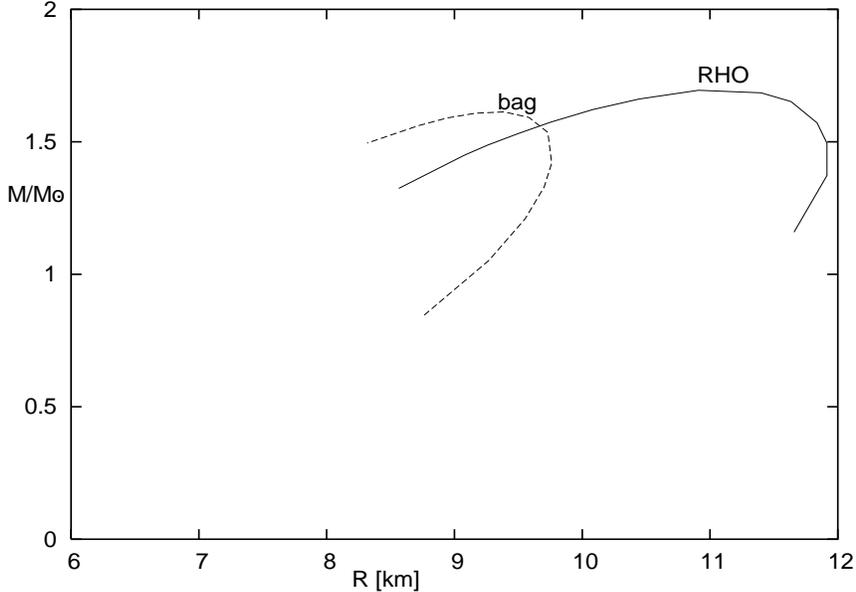}
\caption{ Mass (in terms of solar mass $M_{\odot}$) vs. radius (in km) plot for quark star using RHO model and bag model for $C_q = B^{1/4} =$ 145 MeV and $m_q = $ 145 MeV.} 
\end{figure}

\section{Bag model for quark star:}

For comparison, let us briefly review quark stars in bag model \cite{ra.1}. Here the combined pressure due to gravitation and bag pressure balances the degenerate pressure to form quark star. Using the standard procedure again, the energy, 
\begin{equation}
E_0 = g_f \sum_{p=0}^{p_F} \sqrt{p^2 + m_q^2} + B V = \frac{g_f V}{2 \pi^2} \int_0^{p_F} dp\,p^2\, \sqrt{p^2 + m_q^2} + B V \,\,, \label{eq:eb}
\end{equation}
and the number of quarks,
\begin{equation}
N = g_f \sum_{p=0}^{p_F} = \frac{g_f V}{2 \pi^2} \frac{p_F^3}{3}\,\,,
\end{equation}
where $p_F$ is the Fermi momentum. From the energy, Eq. (\ref{eq:eb}), degenerate pressure may be obtained and using the same TOV equations, Eq. (\ref{eq:tovp}, \ref{eq:tovm}), mass-radius relationship of quark star may be obtained. Now the variables are normalized as,
\begin{equation}
\bar{P} = \frac{2 \pi^2}{g_f B} P\,\,;\,\, \bar{\varepsilon} = \frac{2 \pi^2}{g_f B} \varepsilon\,\,; \,\,\bar{M} = \left( \frac{g_f}{2 \pi^2}\right) ^{1/2} B^{1/2} G^{3/2} M\,\,;\,\,\bar{r} = \left( \frac{g_f}{2 \pi^2}\right) ^{1/2} B^{1/2} G^{1/2} r\,\,. 
\end{equation}
Again TOV is solved numerically for different initial values of central densities to obtain mass-radius relationship, $\bar{M}$ and $\bar{r}$ corresponding to the maximum mass is obtained. From $\bar{M}$ and $\bar{r}$, the actual mass and radius are obtained from,
\begin{equation}
M = \bar{M} \left( \frac{2 \pi^2}{g_f}\right)^{1/2} \frac{1}{B^{1/2} G^{3/2}} \,\,
\end{equation}
\begin{equation}
R = \bar{R} \left( \frac{2 \pi^2}{g_f}\right)^{1/2} \frac{1}{B^{1/2} G^{1/2}} \,\,,
\end{equation}
which has a similar structure as that of quark star in RHO and mass-radius relationship is plotted in Fig. 1. $C_q$ of RHO is replaced here by $B^{1/4}$. Similar mass and radius dependence on bag constant is also demonstrated in Ref.\cite{ra.1} using a simple energy principle. 

\section{Results:}

The TOV equations, Eq. (\ref{eq:tovp}, \ref{eq:tovm}), is numerically solved for both the models with different central densities. We have considered quark star with two flavors and hence $g_f = 12$. The mass of the star increases with the central density and reaches a maximum value and decreases as shown in Fig. 1. There is a clear difference between the two models. The maximum mass and corresponding radius are tabulated in Tables 1 and 2. We have taken $C_q = B^{1/4} = 145\, MeV$, which is in the range of values used in hadron spectroscopy, and obtained the maximum mass for different quark masses, $0$, $145\,MeV$ and $m_p/3$, where $m_p$ is the proton mass. $M_{max}$ in RHO is almost same as that in bag model for massless quarks, but differ for finite masses. $M_{max}$ and corresponding radius both are higher in RHO than bag model. $M_{max}$ in bag model is almost same as that of Ref. \cite{ra.1}. For massless quarks, quark star mass as well as radius scales as $C_q^{-2}$ or $(B^{1/4})^{-2}$, the confinement strengths. Therefore, larger the confinement strength smaller the mass and the radius. From the mass-radius plot, Fig. 1, we see that the stable quark stars in RHO model always have larger mass and radius compared to the bag model. From Tables 1 and 2, the central energy densities required for stable stars are lower in RHO model. For stable stars, mass increases with central energy density, and hence have lower central energy density than that of maximum mass star.        
\begin{table}
\caption{Mass, Radius, energy density and quark density of quark star having maximum mass for different quark masses in RHO model with $C_q$ = 145 MeV.}  
\begin{center}
\begin{tabular}{||c|c|c|c|c||}
\hline
$m_q$ &$M_{max}$ &$R$ &$\varepsilon_c$ &$n_c$  \\  
$MeV$ &$M_{\odot}$ &$km$ &$GeV/fm^3$ & $fm^{-3}$  \\
\hline
0 &1.96 &12.55 &0.87&2.42  \\  
\hline
145 &1.7 &10.9 &1.34&3.22  \\  
\hline
$m_p/3$ &1.14 &8.15 &2.14&4.085 \\  
\hline
\end{tabular}  
\end{center} 
\end{table}

\begin{table}
\caption{Mass, Radius, energy density and quark density of quark star having maximum mass for different quark masses in bag model with $B^{1/4}$ = 145 MeV.}   
\begin{center}
\begin{tabular}{||c|c|c|c|c||}
\hline
$m_q$ &$M_{max}$ &$R$ &$\varepsilon_c$ &$n_c$  \\  
$MeV$ &$M_{\odot}$ &$km$ &$GeV/fm^3$ & $fm^{-3}$  \\
\hline
0 &1.98 &11.22 &1.05&2.775  \\  
\hline
145 &1.6 &9.4 &1.41&3.31  \\  
\hline
$m_p/3$ &1.02 &6.55 &2.98&5.32 \\  
\hline
\end{tabular}  
\end{center}
\end{table} 

\section{Conclusions:}
  
Here we have applied the RHO model of hadrons to study quark star, using TOV equations and compared with extensively studied bag model. In RHO model, confinement leads to discrete Landau levels and associated Landau degeneracy, like in the problem of Landau's diamagnetism, instead of continuum levels as in the case of bag model. Hence, the EoS is different from that of bag model and mass-radius relationship is obtained for quark star using this new EoS and TOV. On comparison with that of bag model, there is a clear difference in mass-radius relationship. For finite quark masses the maximum mass of quark star and corresponding radius are higher in RHO than that of bag model because of lower average confinement pressure inwards. Masses and radii of stable quark stars are higher, but central energy densities are lower in RHO model than that of bag model.

\end{document}